\documentstyle[epsfig,ifthen,multicol,aps,prl]{revtex}

\begin{document}
\draft 

\title{Formation of an Edge Striped Phase in Fractional Quantum Hall Systems}

\author{E.~V.~Tsiper and V.~J.~Goldman}
\address{Department of Physics, State University of New York, Stony Brook, NY 11794-3800}

\date{December 21, 2000}

\maketitle

\begin{abstract}
We have performed an exact diagonalization study of up to $N=12$ interacting electrons on a disk at filling $\nu=\frac{1}{3}$ for both Coulomb and $V_1$ short-range interaction for which Laughlin wave function is the exact solution. For Coulomb interaction and $N\geq 10$ we find persistent radial oscillations in electron density, which are not captured by the Laughlin wave function. Our results srongly suggest formation of a chiral edge striped phase in quantum Hall systems. The amplitude of the charge density oscillations decays slowly, perhaps as a square root of the distance from the edge; thus the spectrum of edge excitations is likely to be affected.
\end{abstract}

\vspace{-0.1cm}
\begin{multicols}{2}

An important open problem in the physics of the fractional quantum Hall effect (FQHE) \cite{Tsui82,Laughlin83} is the structure and the properties of edge states. The bulk FQHE is reasonably well understood: the kinetic energy of 2D electrons is quenched by the strong perpendicular magnetic field $B$, and the Coulomb interaction dominates the physics of the partially filled Landau level (LL). At certain filling factors $\nu$ the electrons condense into highly correlated gapped quantum fluids, which results in quantization of the Hall conductance $\sigma_{xy}$ and vanishing diagonal conductivity $\sigma_{xx}$ \cite{Books}. There has been much interest in the physics of FQH edge states for several reasons. First, on a QH plateau the bulk electron states are localized by disorder, and gapless excitations are possible only in extended edge states \cite{Lgauge,Hedge}; thus the transport current flows along the edges \cite{JKWang}. Second, FQH edge states have played an important role in many experiments: internal resonant tunneling from one edge to another through states bound on a quantum antidot has been employed to measure the fractionally quantized charge of the tunneling quasiparticles \cite{SciCharge}, and external electron tunneling into an edge of a FQH system has been used to study its excitation spectrum \cite{Chang96}. Also, an edge channel is usually assumed to be localized within a few magnetic lengths $\ell_0 =\sqrt {\hbar c/eB}$ at the boundary of the FQH system; then the low-energy dynamics is effectively 1D, and field-theoretic descriptions of edge channels as a chiral Tomonaga-Luttinger liquid have been developed \cite{Wen,Stone}.

Our quantitative understanding of the FQHE is based on Laughlin wave function \cite{Laughlin83}, which is known to be quite accurate for the bulk, translationally invariant $\nu=\frac{1}{3}$ FQH state from extensive comparisons with exact diagonalization results in the edgeless spherical geometry \cite{Haldane83,SHe}. However, the accuracy of this wave function remains relatively untested at the edge. Validity of a wave function in the bulk does not necessarily extend to the edge: the ground state of the system is separated from excited states by a gap in the bulk, which makes it insensitive to perturbations; on the other hand, the system has gapless excitations at the edge, which makes it more susceptible to the particular form of the interelectron interaction. The FQHE has been relatively less studied in the disk geometry \cite{Laughlin83,Dev92,Cappelli}, because, in part, the interest was focused on the bulk states, and it is difficult to separate disk edge effects for small number of electrons.

In this paper we report results of a detailed numerical investigation of the microscopic structure of the FQH edge. To this end, we diagonalize the interaction Hamiltonian in the disk geometry for up to $N=12$ Coulomb-interacting electrons. For $N\geq10$ we observe formation of a striped order in the charge density at the edges. These edge stripes are not captured by the Laughlin wave function that works well for the bulk FQH states. We also obtain an analytical fit to the numerical data, which suggests that the amplitude of the charge density oscillations decays only as a power law (specifically, as inverse square root) with distance, appreciably extending into the sample. We interpret these results as formation of an edge striped phase (ESP) with wave vector $q_{\sc esp}\approx \pi/2\ell_0$, possibly smectic liquid crystal, at the edge of a FQH system.

We study the simplest FQH state, that of spin-polarized electrons restricted to lowest Landau level, as appropriate in the large $B$ limit, at filling $\nu=\frac{1}{3}$. The interaction Hamiltonian in the second-quantized form is

\begin{equation}
{\cal H}=\sum_{mnl}V_{mn}^l c_{m+l}^\dagger c_n^\dagger c_{n+l}c_m ~,
\end{equation}

\noindent    where $c_m^\dagger$ creates an electron in the single-particle eigenstate of the angular momentum $\left|m\right >$, where $m=0$,1, $\ldots$. There is no external confinement; the Coulomb matrix elements $V_{mn}^l$ can be expressed via finite sums \cite{preprint,GirvJach}. The bulk ground state of $N$ interacting electrons is well approximated by the Laughlin wave function: 

\begin{equation}
\Psi_{\rm L}(z_1, ...\, , z_{\rm N}) = \prod_{j<k}^N(z_j-z_k)^3\exp\{\,-\sum_j^N \left|\frac{z_j}{2}\right|^2\} ~,
\label{PsiL}
\end{equation}

\noindent   where $z_j=x_j-iy_j$ (in units of $\ell_0$) gives the position of $j$-th electron as a complex number. This wave function is known to be the exact ground state \cite{Haldane83,TruKiv} of the electrons interacting via short-range potential $V_1$ defined by expressing the Coulomb interaction as 

\begin{equation}
V(|z_j-z_k|)=\sum_l V_l P_l^{jk}  ~,
\label{Vl}
\end{equation}

\vspace{-0.3cm}
\noindent   where $V_l$ are the Haldane pseudopotentials, and $P_l^{jk}$ is the projection operator which selects the states in which particles $j$ and $k$ have relative angular momentum $l$ \cite{Haldane86}. 

We construct $\Psi_{\rm L}$ and $\Psi_{\rm C}$ numerically as the ground states of the short-range and Coulomb Hamiltonians, respectively, for the total angular momentum $M=\frac{3}{2}N(N-1)$ for $N$ electrons, which gives filling $\nu=\frac{1}{3}$ in thermodynamic limit $N\rightarrow \infty$. The Hilbert space is restricted by consideration of single-particle orbitals with angular momentum $m\leq m_{\rm max}$ only. For $V_1$ short-range interaction, the contribution of orbitals with $m>m_{\rm max}^{\rm L}=3(N-1)$ vanishes identically; for Coulomb interaction $m_{\rm max}^{\rm C}$ is obtained by increasing Hilbert space untill overlap $\left<\Psi_{\rm L}|\Psi_{\rm C}\right>$ converges to at least three significant digits. For example, for $N=12$, the largest $\nu=\frac{1}{3}$ FQH system studied, $M=198$ and the size of the Hilbert space is 15,293,119 for $m_{\rm max}^{\rm C}=35$. To the best of our knowledge, the largest $\nu=\frac{1}{3}$ systems studied in the disk geometry prior to this study was $N=10$ by Cappelli {\em et al.}\ \cite{Cappelli}, who used only $V_1$ interaction.

In Fig.\ 1 we present the radial electron density profiles $\rho_{\rm L}$ and $\rho_{\rm C}$ for both $\Psi_{\rm L}$ (short-range interaction) and $\Psi_{\rm C}$ (true Coulomb interaction) for $N=5$ to 12 electrons. A comparison of $\rho_{\rm L}(r)$ with the exact Coulomb $\rho_{\rm C}(r)$ should clarify the importance of the long-range part of the Coulomb interaction for the FQH edge structure. For all $N$, the exponential fall-off of $\rho_{\rm C}(r)$ at the edge and the density oscillation nearest the edge is captured well by the Laughlin $\rho_{\rm L}(r)$. As expected, the edge shifts to larger $r$ in accordance with $r_{\rm edge}\approx (2m_{\rm max}^{\rm L})^{1/2}$. The properly normalized overlaps $\left<\Psi_{\rm L}|\,\Psi_{\rm C}\right>$ are given in Table I; the corresponding overlaps are much closer to unity in the spherical geometry, which indicates that the deviation originates in the disk edge. Still, given the huge size of the Hilbert space, the overlaps indicate that Laughlin $\Psi_{\rm L}$ capture the important short-range FQH correlations remarkably well, even at the edge. Table I also gives $W$, the weight of the components of $\Psi_{\rm C}$ with $m> m^{\rm L}_{\rm max}$. All $W<1\%$, which means that fixing total angular momentum $M=\frac{3}{2}N(N-1)$ selects the $\nu=\frac{1}{3}$ state for Coulomb interaction too: it fixes {\em average} density on the disk $\left<\rho_{\rm C}\right>\approx \left<\rho_{\rm L}\right>$ (for $r<r_{\rm edge}$) to better than $10^{-3}$.

For $N=6$ to 8, density profile $\rho_{\rm C}$ clearly displays a commensurability effect, strongest for $N=7$, which can be visualized in a classical picture as tendency to have one central electron at $r=0$ surrounded by a ring of six electrons. For $N=7$ the position of the outer maximum in the density profile is very close to the lattice constant $a_{\sc wc}=(4\pi \sqrt {3})^{1/2}\ell_0\cong 4.665\ell_0$, of the $\nu=\frac{1}{3}$ 2D Wigner crystal (WC) in the thermodynamic limit \cite{WC}. This commensurability effect is barely visible for the Laughlin density profile $\rho_{\rm L}(r)$, indicating importance of the long-range part of Coulomb interaction for formation of an ordered phase, even for such small electron systems.

\begin{figure}
\renewcommand{\baselinestretch}{0.85}
\centerline{\epsfig{file=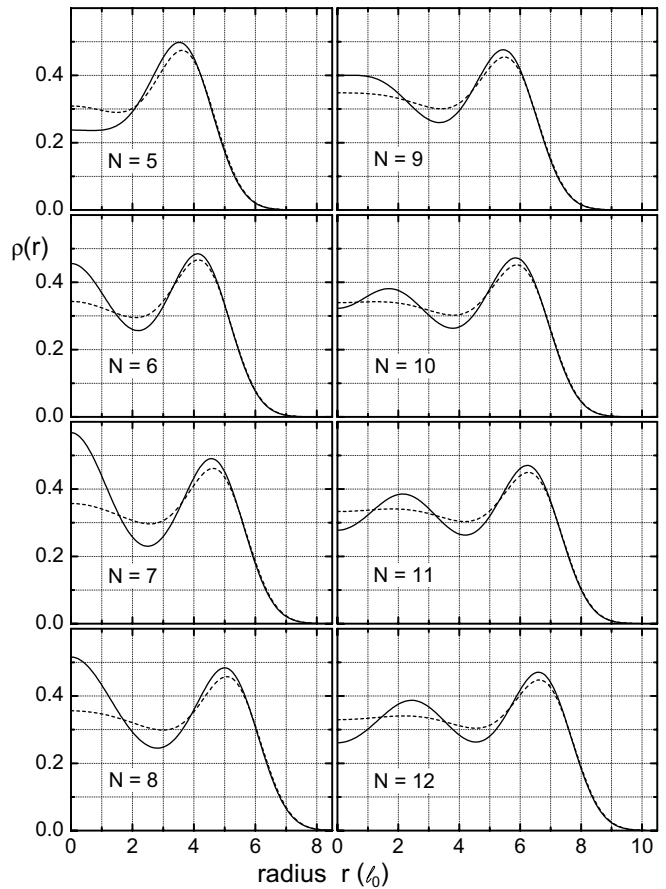, width=8.6cm}}    
\vspace{0.2cm} 
\caption{Electron densities $\rho_{\rm L}(r)$ and $\rho_{\rm C}(r)$ for $N$ interacting electrons in the disk geometry. The dashed lines show the Laughlin state for a short-range interaction, solid lines show the exact ground state for the Coulomb interaction.  }
\label{FIG1}
\end{figure}

\begin{table}[h]
\renewcommand{\baselinestretch}{0.98}
\caption{ Normalized overlaps of the Laughlin ground state $\Psi_{\rm L}$ with the exact Coulomb ground state $\Psi_{\rm C}$, and the Coulomb ground state in the restricted Hilbert space $\Psi_{\rm Cr}$ for $N$ electrons. For $\Psi_{\rm Cr}$ the single-particle angular momentum $m$ is restricted to $m\leq m^{\rm L}_{\rm max}$, where $m^{\rm L}_{\rm max}$ is for $\Psi_{\rm L}$. Here $W$ is the weight of the components in the exact $\Psi_{\rm C}$ with $m> m^{\rm L}_{\rm max}$. \\
$ ^a$ For $N=12$ ``unrestricted" overlap was obtained by an extrapolation, using the results for $m^{\rm C}_{\rm max}\leq m^{\rm L}_{\rm max}+2$ .  }
\vspace{0.3cm} 
\renewcommand{\arraystretch}{1.1}
	\begin{tabular}{|c|l|l|l|}
	\renewcommand{\arraystretch}{1.1}
$~N~$ & $\left<\Psi_{\rm L}|\,\Psi_{\rm Cr}\right>~~~~$ &  $\left<\Psi_{\rm L}|\,\Psi_{\rm C}\right>~~~$  &  $10^3\times W~~~~$ \\[1pt]
	\tableline
 3  & 0.993606  &  0.990997      &  3.81  \\
 4  & 0.983108  &  0.978804      &  5.20  \\
 5  & 0.988168  &  0.985047      &  4.10  \\
 6  & 0.986102  &  0.981767      &  4.61  \\
 7  & 0.964758  &  0.956392      &  6.92  \\
 8  & 0.968652  &  0.962187      &  6.49  \\
 9  & 0.972992  &  0.966526      &  6.61  \\
 10 & 0.971415  &  0.964413      &  7.08  \\
 11 & 0.966118  &  0.958138      &  7.67  \\
 12 & 0.961240  &  0.9528$^a$    &  7.7$^a$ \\
	\end{tabular}  
	\renewcommand{\arraystretch}{1.0}
\label{Table1}
\end{table}  

\noindent    

For $N\geq10$ the Laughlin state $\Psi_{\rm L}$ develops a constant density plateau $\rho_{\rm L}\approx\frac{1}{3}$ in the interior of the disk. This is the precursor of the translationally and rotationally invariant $\nu=\frac{1}{3}$ bulk FQH condensate. Surprisingly, a second oscillation in the Coulomb $\rho_{\rm C} (r)$ becomes evident for $N=10$; the second oscillation shifts to larger $r$ as $N$ is increased, following the first oscillation. The second oscillation is clearly caused by the long-range part of the Coulomb interaction, in other words, by the Haldane pseudopotentials $V_l$ with $l\geq 3$ (because of Fermi statistics, $l$ must be odd for spin-polarized electrons). In view of large overlaps, one may want to describe the edge charge density oscillations in $\Psi_{\rm C}$ in terms of excitations of $\Psi_{\rm L}$ by the long range part of the Coulomb interaction. Since the charge and the angular momentum $M$ of the total system are fixed, a relevant elementary excitation to consider is a quasiparticle-quasihole bound exciton \cite{Lexciton}. The excitations that do not conserve $m$ are not relevant because they alter $M$. A variational wave function that admixes one bound exciton into $\Psi_{\rm L}$ reproduces charge density oscillations in $\rho_{\rm C}(r)$ for $N=10$ to 12 reasonably well \cite{private}. The deviation of the variational $\rho (r)$ from the exact $\rho_{\rm C}(r)$ increases, however, as $N$ is increased from 10 to 11 to 12, which indicates that more than one bound exciton is needed for larger systems.

We have obtained a good phenomenological fit of the exact $\rho_{\rm C}(r)$ dependence using an analytical expression:

\begin{equation}
\rho(r)=\rho_0[(\mbox{erf}\, \bar{r}+1)/2]\{1+\rho_{\sc esp0}{\rm J}_0[\,q_{\sc esp}(\bar{r}-\bar{r}_0)]\}\,,
\label{fit}
\end{equation}

\noindent     where $\rho_0=\frac{1}{3}$ for $\nu=\frac{1}{3}$, $\bar{r}=r_{\rm edge}-r$ is the distance from the edge in units of $\ell_0$, $\mbox{erf}(x)$ is the Gauss error function, and ${\rm J}_0$ is the Bessel function of the first kind. The fitting parameters are the ESP oscillation amplitude at the edge $\rho_{\sc esp0}$, the ordering wave vector $q_{\sc esp}$, and the shift of the first oscillation from the edge $\bar{r}_0$. The fit is robust in the sense that the same fitting parameters also give good fits for $N=10$ and 11, scaling $r_{\rm edge}$ as $(2m_{\rm max}^{\rm L})^{1/2}$. A fit for $N=11,12$ is shown in Fig.\ 2, where the values of the parameters are: $\rho_{\sc esp0}=0.519$, wave vector $q_{\sc esp}=1.570/\ell_0$, and $\bar{r}_0=0.99$. We believe that the deviation of the fit from the exact density in the center of the disk for $r\!<\!1$ is a finite-size effect. The parameters $\rho_{\sc esp0}$ and $\bar{r}$ may be varied by 1\% to give equally good overall fit. The wavelength $\lambda_{\sc esp}=4.002$ can be compared to several natural lengths in the problem. It is significantly smaller than the Wigner crystal lattice constant $a_{\sc wc}=(4\pi \sqrt {3})^{1/2} \cong 4.665$, and, more relevant, is accurately equal to the distance between the crystal planes $h_{\sc wc}=(\sqrt {3}/2)a_{\sc wc}=(3\pi \sqrt {3})^{1/2}\cong 4.040$. We also note that $q_{\sc esp}$ is close to the wave vector of the bulk mag\-ne\-to\-roton minimum $q_{\sc mr}$ \cite{GMP}.  Similar length scales are provided by many  other quantities, {\em e.g.}: the ``ion diameter" $2R_0=\sqrt {6\pi}\cong 4.342$, a little too large; square root of the ``disk area per electron" $(2\pi m_{\rm max}^{\rm L}/N)^{1/2}=[6\pi (N-1)/N]^{1/2}\cong 4.157$ for $N\!=\!12$ is closer, but the $\rho_{\rm C}(r)$ data do not show such dependence on $N$, Fig.\ 2.

\begin{figure}
\renewcommand{\baselinestretch}{0.85}
\centerline{\epsfig{file=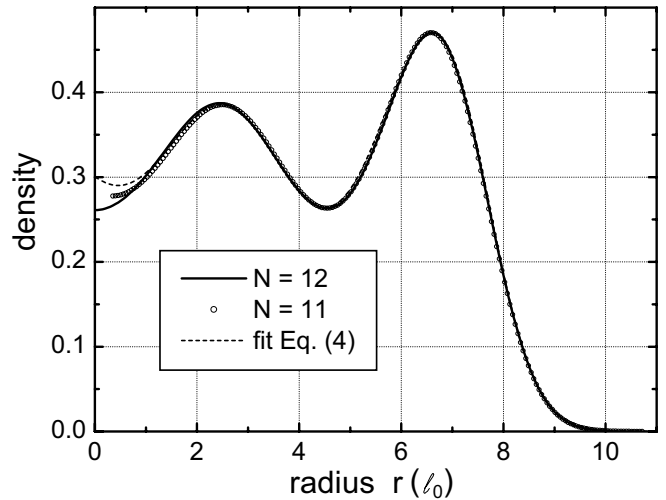, width=8.6cm}}   
\vspace{0.2cm} 
\caption{ Electron density $\rho_{\rm C}(r)$ for $N$ Coulomb-interacting electrons. The $N=11$ density data (circles) is shifted radially by $r_{\rm edge}(N=12)-r_{\rm edge}(N=11)\approx0.378$. The dashed line is the fit of Eq.\ (4) with parameters given in the text. }
\label{FIG2}
\end{figure}

In the disk geometry rotational symmetry (corresponding to translational invariance {\em along} a linear edge) of the chiral basis orbitals $\left|m\right >$ ensures that azimuthal density modulations in $\rho(\theta )$ are not possible.  A broken symmetry ground state is possible only as a superposition of     several rotationally invariant {\em degenerate} ground states at different total $M$. Such ground state would be pinned by disorder and QHE would not be observable in experiments. A charge density wave (CDW) order is possible in the radial direction, however; in the thermodynamic limit \cite{limit}, the edge confinement breaks the translational invariance and ``excites" radial CDW. Because of the long-range nature of the Coulomb interaction, the radial CDW is expected to propagate deep into the interior of the electron system and form a locally anisotropic edge striped phase (ESP). Since there is abundant experimental evidence that FQH edges are conducting, ESP is a conductor along the edge, and an insulator in the perpendicular direction, in the sense that in a QH system radial perturbation will induce only azimuthal Hall current. Thus, ESP can be thought of as a quantum smectic liquid crystal phase \cite{EmeryK} at the edge of the system.

Extrapolating to the thermodynamic limit, and setting $\rho_{\sc esp0}=\frac{1}{2}$, $q_{\sc esp}=\frac{\pi}{2}$, the following simple expression describes well the electron density at the $\nu=\frac{1}{3}$ FQH edge:

\begin{equation}
\rho(y)=\frac{1}{6}(\mbox{erf}\, y+1)\{1+\frac{1}{2}\,{\rm J}_0[\,\frac{\pi}{2}(y-1)]\}  ~,
\label{rho}
\end{equation}
\noindent    where $y$ is the distance from the edge. In the disk geometry the azimuthal transport current density $J_{\theta }\propto \partial_r \rho (r)$. Differentiating Eq.\ 6 we obtain edge current density

\begin{eqnarray}
J_{x}(y)\propto \frac{1}{3\sqrt{\pi}}(e^{-y^2})\{1+\frac{1}{2}\,{\rm J}_0[\,\frac{\pi}{2}(y-1)]\} \nonumber\\
 - \: \frac{\pi}{24}(\mbox{erf}\, y+1)\,{\rm J}_1[\,\frac{\pi}{2}(y-1)] ~.
\label{current}
\end{eqnarray}

\begin{figure}
\renewcommand{\baselinestretch}{0.85}
\centerline{\epsfig{file=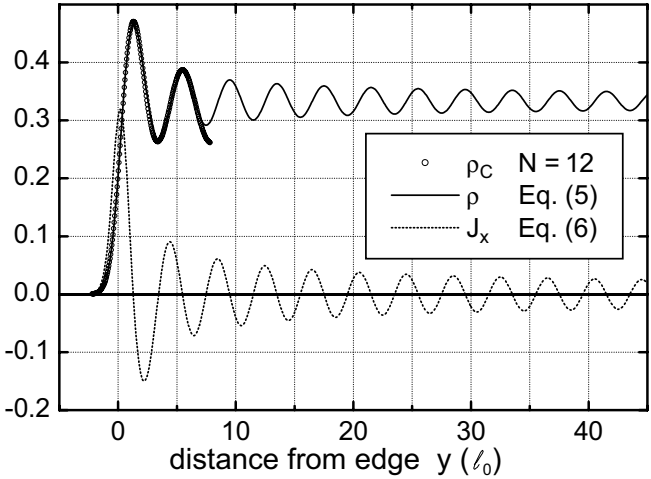, width=8.6cm}}   
\vspace{0.2cm} 
\caption{ Electron density and transport current density (arb. units) {\em vs.}\ distance from the $\nu=\frac{1}{3}$ FQH edge. Density $\rho_{\rm C}(r)$ for 12 electrons is also shown. The analytical expressions Eqs.\ 5 and 6 are inferred from the numerical data.  }
\label{FIG3}
\end{figure}

\noindent     In practice, the first term with $\exp(- y^{2})$ vanishes after the first oscillation. The phenomenological expressions Eqs.\ 5 and 6 are plotted in Fig.\ 3. Large $y$ asymptotic     behavior of both Bessel functions is $\propto y^{-1/2}\cos (\pi y/2)$ (with a $\frac{\pi}{2}$ phase shift). Thus, should Eqs.\ 5 and 6  prove  applicable to linear edges in large systems, the amplitudes of the charge density and the transport current oscillations both decay only as a power law of distance from the edge, specifically as $y^{-1/2}$. We note that the $\propto y^{-1/2}$ asymptotic behavior is not well established by our numerical results. 

There are likely to be several important implications of the observed slow-decaying charge density oscillations at FQH edges. Note first that they do not affect topological QHE properties, such as the values of the quantized $\sigma_{xy}$ \cite{Tsui82} and the charge of the bulk Laughlin quasiparticles \cite{SciCharge}. On the other hand, dynamic properties, where edge excitations at finite frequency and wave vector are important, are likely to be affected. For example, since edge excitations are expected to be transverse charge density waves \cite{Wen}, it seems that the presence of ESP at QH edges should affect spectrum of edge excitations \cite{Chang96}. The striped phase may be the reason for the reported experimental dependence of velocity of propagation of edge magnetoplasmons on frequency \cite{EMPlasmon}. When there are several edge modes present, a striped phase extending hundreds of $\ell_0$ from the edge is likely to couple various modes, and this may explain the failure to observe multiple and/or counterpropagating modes in experiments \cite{EMPlasmon}.

A complementary field-theoretic approach for the dynamics of an edge channel treats it as a Luttinger liquid, focusing on the asymptotic long-distance, low-energy physics \cite{Wen,Stone}. It has been argued that many expected power law behaviors in this limit are completely determined by the quantized value of $\sigma_{xy}$, and would not be affected by the specifics of the edge structure. The mapping of a FQH edge onto the Luttinger liquid is based on rather general considerations, {\em provided} there is no $\propto1/r$ interparticle interaction. In this work we specifically show that the long-ranged true Coulomb interaction leads to formation of a striped phase at the $\nu=\frac{1}{3}$ FQH edge, calling into question the fundamental assumption that all important physics at the edge is one-dimensional.

We are grateful to I.\,L.\,Aleiner and J.\,K.\,Jain for discussions. This work was supported in part by the NSF under Grant No. DMR9986688.

\bibliographystyle{prsty}

\begin{references}
\bibitem{WC}  WC can't be the ground sate at $\nu\!=\!\frac{1}{3}$ without LL mixing.
\bibitem{limit} In the thermodynamic limit  azimuthal angle $\theta $ degree of freedom maps onto coordinate $x$ along the  linear edge, and the radius $r$ maps onto $y$ transverse to the edge. We also note that confinement by fixing $M$ is very sharp; electron confinement by a shallow external potential $V_{\rm ext}$ (small gradient: $\ell_0\partial_r V_{\rm ext}\ll \Delta_{\rm gap}$) may reduce amplitude of the ESP, but can also lead to phase separation.
\end{references}

\end{multicols}

\end{document}